# Using an Epidemiological Model to Study the Spread of Misinformation during the Black Lives Matter Movement

[1] Maryam Maleki, Esther Mead, Mohammad Arani, and Nitin Agarwal

*Abstract*— The proliferation of social media platforms like Twitter has heightened the consequences of the spread of misinformation. To understand and model the spread of misinformation, in this paper, we leveraged the SEIZ (Susceptible, Exposed, Infected, Skeptics) epidemiological model to describe the underlying process that delineates the spread of misinformation on Twitter. Compared to the other epidemiological models, this model produces broader results because it includes the additional Skeptics (Z) compartment, wherein a user may be Exposed to an item of misinformation but not engage in any reaction to it, and the additional Exposed (E) compartment, wherein the user may need some time before deciding to spread a misinformation item. We analyzed misinformation regarding the unrest in Washington, D.C. in the month of March 2020 which was propagated by the use of the #DCblackout hashtag by different users across the U.S. on Twitter. Our analysis shows that misinformation can be modeled using the concept of epidemiology. To the best of our knowledge, this research is the first to attempt to apply the SEIZ epidemiological model to the spread of a specific item of *misinformation*, which is a category distinct from that of *rumor*, and *hoax* on online social media platforms. Applying a mathematical model can help to understand the trends and dynamics of the spread of misinformation on Twitter and ultimately help to develop techniques to quickly identify and control it.

*Keywords*— Black Lives Matter, Epidemiological model, Mathematical modeling, Misinformation, SEIZ model, Twitter,

## I. INTRODUCTION

IN recent years, social media has become one of the most important sources of information for information consumers all over the world, especially for the consumption of news [1], [2]. According to a Pew Research report, the number of Americans who obtain their news through social media instead of print newspapers has drastically increased [3]. Due to its popularity, accessibility, low-barrier for publication, and crowd-sourced nature, Twitter is faced with the risk of the rapid dissemination of misinformation. The ability to post information on Twitter immediately after an event occurs, or even as it is occurring, causes the reliability of the news to potentially be questionable [4], [5]. For example, a considerable number of Americans were Exposed to misinformation prior to the 2016 U.S. presidential election, and some surveys indicate that many users who were Exposed to false narratives decided to believe them [6]. In addition, some pundits have argued that misinformation had a significant effect on the actual outcome of the 2016 U.S. presidential election [7].

Wu et al. defines *misinformation* as inaccurate information that is unintentionally propagated by users, while *disinformation* is fake information that is intentionally developed and propagated to mislead people. According to this definition, the key difference between misinformation and disinformation lies in the *purpose* of its propagation [8].

Similarly, Shu et al. state that *misinformation* is false information spread by a user who does not know it is inaccurate or misleading. Also, similarly, these researchers define *disinformation* as false or inaccurate content that is deliberately propagated to deceive people [9]. Since any user can post information items on social media, it is not easy for researchers to determine whether an item of misinformation is intentionally created or not [9]. Therefore, in this research we make no assumptions upon this point, and, instead, focus on the overall concept of misinformation.

Shu et al. also define a *rumor* as a story spread from individual to individual, wherein the truth of the specific information item is not verified or is otherwise considered to be dubious. Rumors usually emerge in the existence of vague or threatening incidents. When it is proved that a rumor is false, it then becomes a type of *misinformation* [9]. For the purposes of this study, it is important to differentiate between these concepts of rumor, disinformation, and misinformation to ensure that they are clearly delineated.

This work is motivated by the power that misinformation can have in the minds of social media consumers, especially relative to politically sensitive issues. We attempted to apply a mathematical model to explain how a specific piece of politically charged misinformation was able to be quickly propagated across Twitter even though it was entirely false.

We specifically evaluated the dissemination of misinformation about the extent of the unrest in Washington, D.C., the capital of the United States. This misinformation item claimed that there was a communication blackout in D.C. because of riots supposedly occurring on Monday, June 1, 2020. We were motivated to use a specific epidemiological model to study how this misinformation item diffused on Twitter.

[1]M., Maleki, University of Arkansas at Little Rock, Little Rock, AR, 72204 USA (mmaleki@ualr.edu).
E., Mead, University of Arkansas at Little Rock, Little Rock, AR 72204 USA (elmead@ualr.edu)

A., Arani, University of Arkansas at Little Rock, Little Rock, AR 72204 USA (mxarani@ualr.edu
N., Agarwal, University of Arkansas at Little Rock, Little Rock, AR 72204 USA (nxagarwal@ualr.edu).

Epidemiological models divide the population into different compartments that represent the state of each user involved in the social network. In this study, we have used the SEIZ (Susceptible, Exposed, Infected, and Skeptic) model [10].

Unlike the traditional epidemiological models such as SIS and SIR, our chosen model has the additional Exposed compartment (E), which consists of individuals who need some time before they become Infected by the misinformation, which is indicative of their having decided to react to it by either commenting on it or spreading it. Furthermore, our chosen model has an additional Skeptic (Z) compartment, which consists of users who have heard about the misinformation item but decided not to engage in any reaction to it.

To the best of our knowledge, there has been no prior research that has attempted to apply the SEIZ epidemiological model to the spread of a particular item of *misinformation* on a social media platform. As we mentioned earlier, rumor is different from misinformation, and, although the SEIZ model has been applied to rumor propagation, it has not been applied to misinformation propagation. The objective of this research is to find a mathematical model that represents the spread of misinformation on Twitter.

A mathematical model for the propagation of misinformation allows for the evaluation of the number of people in any compartment at any time, especially the infected compartment, which is the most important compartment as it is composed of the people who actually spread the misinformation.

The remainder of this paper is organized as follows. Section 2 presents the related work that has been done regarding epidemiological modeling for the spread of news and rumors and misinformation. In section 3, the methodology used in our research is described as well as our data collection process. We also briefly discuss the basics of the more traditional epidemic models (SI, SIS, SIR, etc.). These models are then compared to the SEIZ model, which is used in this study. Section 4 discusses the overarching themes and impact of our research. Finally, section 5 concludes the paper with ideas for future work.

## II. RELATED WORKS

By proposing a mathematical model towards the idea of the propagation of contagion in a population on an Online Social Network (OSN), we can acquire potentially useful information about its spread, much like a disease. Consequently, we are able to set the stage for the subsequent implementation of beneficial approaches to control this dissemination [11]. The foundation and mathematical framework of the epidemiological model involves partitioning the total population into various compartments or components.

The SI (Susceptible-Infected) model is the most primitive epidemic model, which divides the population into two groups based on disease status. The Infected component consists of people who are already carrying the disease, and the Susceptible component consists of individuals who do not yet have the disease but are in danger of being Infected by coming into contact with Infected people [12]. Furthermore, individuals who are in the Infected group and have Recovered from the disease may become Susceptible again, which is evaluated in the SIS (Susceptible-Infected-Susceptible) model [11].

Another epidemic model, which is broadly used in different studies is the SIR (Susceptible-Infected-Recovered) model, which has the addition of a Recovered component that consists of people who have acquired immunity to the disease [13], [14], [15].

Khelil et al. developed an epidemic model for information propagation in mobile ad hoc networks (MANET), wherein they evaluated the influence of node density on information propagation [16]. Abdullah et al. applied the SIR (Susceptible, Infected, and Recovered) model for news propagation on Twitter. Their fundamental hypothesis was that there is similarity between the spread of disease and the spread of information propagation on Twitter [14]. The resemblance between disease and rumor propagation in mathematical terms was first studied by Daley and Kendal [17], [18].

Over the years, different models derived from the SIR model were used to evaluate the spread of information and rumors in a population. In one study, Bettencourt et al. applied different epidemiological models containing SIR, SEI, and SEIZ to study the spread of general ideas in a population [10]. In another study, Newman et al. showed that a large class of typical epidemiological models can be solved on a broad variety of networks. They confirmed the accuracy of their results with numerical simulation of SIR epidemics on networks [19].

Kimura et al. applied the SIS model to evaluate the information propagation on a social network where nodes have the possibility to be activated several times [20]. Xiong et al. used a diffusion model (SCIR) containing four statuses: Susceptible, Contacted, Infected, and Refractory (a "state in which nodes cannot be Infected"). These were used to study the information spread on online microblogs. In their model, they highlighted the Contacted status, containing individuals that knew about the information but chose not to spread them. Their research showed that the Contacted individuals could become Infected or Refractory, both of which are stable states [21].

In another study by Jin et al., an epidemiological model was used to specifically evaluate the spread of *news* and *rumors* on Twitter [11]. The authors used the SEIZ (Susceptible- Exposed-Infected-Skeptic) model to characterize information diffusion on social media. In their model, a Skeptic (Z) person was an individual who knew about the news but decided not to engage in any reaction to it, and an Exposed (E) user was a person who had heard about the news but needed some time before deciding to engage in any action [11].

In another study, Wang et al. applied "a variant epidemic model" for rumor propagation on OSNs [15]. They showed that each individual in a network can have one of three states containing the concept of 'Credulous', which is similar to the susceptible group in the SIR model. In their model, a Spreader is an individual who likes to share rumor items with others, and Rationals are similar to the Recovered group in the SIR model. Cheng et al. proposed a stochastic epidemic model for the spread of rumors among social media [22]. In their model, the infectious probability is not a constant, and instead, is a function that depends on the strengths of the ties between "ignorants" and "spreaders".

Tambuscio et al. proposed a modeling framework to evaluate the spread of *hoaxes* on social media, especially how the accessibility of debunking information can potentially contain that spread [23]. Their model can be explained as an SIS model where the Infected state is divided into two components, namely, "believers" and "nonbelievers" [23].

Subsequent research by Tambuscio et al. applied a model to the propagation of the belief in a hoax and the related fact checking in a social network. In their model, an agent, which is an individual, can be a Susceptible (S), "if they have not heard about neither the hoax nor the fact checking, or if they have forgotten about it", or a Believer (B), "if they believe in the hoax and choose to propagate it", and a Fact checker (F), "if they know the hoax is false, for example after having consulted an accurate news source, and chose to spread the fact-checking" [24].

In another study, Dubravka et al. proposed a novel approach wherein they applied a "fuzzy model" to combat the propagation of misinformation on Twitter. In their model, they defined an "influence power", which is the number of various characteristics of an individual that can result in others deciding to believe the information they share on Twitter. In other words, if the "influence power number" of a misinformation item is a very small number, there is not a significant possibility of a prosperous share of that misinformation item [25].

In another study, Cho et al. developed a version of the epidemiological SIR model to evaluate the number of users who believe in different kinds of information such as *true* or *false* information. They evaluated the user's decision to believe accurate or inaccurate information using an "estimated uncertainty" in their expected belief or doubt for propagating misinformation on social media [26].

In another paper, Cho et al. suggested a Subjective Logic (SL) model to evaluate how to destroy or reduce the influence of misinformation by spreading true information against the misinformation. They map users' belief components into each compartment in the SIR epidemiological model to evaluate the ratio of Recovered users who believe in accurate information. Their findings demonstrate that users' prior opinion about misinformation can influence their decisions toward a belief in accurate information even if there is a high uncertainty situation [27].

Although the previous literature is quite comprehensive, our work stands out in that it applies one of the most robust epidemiological models (SEIZ) to a very specific item of known misinformation on Twitter, which caused a significant level of disruption on the social media platform, quickly becoming a "trending" hashtag wherein many of the tweets were attempting to mobilize individuals to become physically active and even to incite violence [30]. Next, we present our research methodology.

III. METHODOLOGY

Our methodology starts with data collection followed by the application of the SEIZ model. We first outline the components of the basic epidemic models to provide a background and comparison. Then we elaborate on the target SEIZ model and the process of "learning" its parameters. We used the SEIZ model because it includes the Skeptics component, wherein a user may be exposed to misinformation but decide not to engage in any reaction to it. It also has the Exposed compartment wherein the user can spend some deliberative time before making any decision regarding spreading the misinformation [10], [11].

*A. Data Collection*

The focus of this study was to analyze a specific misinformation item regarding the civil unrest that happened in Washington, D.C. in the month of March 2020. This event was a culmination of the "Black lives Matters" protest that sprouted due to the death of George Floyd [31]. As a result, there was a widespread misinformation campaign regarding a communication outage that supposedly happened on Monday, June 1, 2020 in D.C.

This was propagated by the use of the #DCblackout hashtag by various actors across the US on Twitter. For our study, we extracted 27,962 tweets that used the #DCblackout hashtag from June 1-4, 2020 using Python's "Twint" library [32]. Numerous Twitter accounts shared photos of a large, out-of-control fire close to the Washington Monument. However, some other accounts mentioned that the photos were copied from the TV show "Designated Survivor" and were not related to the riot and unrest at all [33].

*B. Model*

As mentioned earlier, we used an epidemiological model, which divides the population into different compartments or components to study the spread of misinformation on social media, and specifically Twitter. We now describe two different preliminary epidemiological models, SIS and SIR, and compare them with the SEIZ model which is used in this study.

*C. SIS Model*

The SIS model divides the population into two different compartments: Susceptible (S) and Infected (I) (Fig. 1). As there is no accounting for any immunity against the disease in this model, the Infected person returns to the Susceptible component. To adjust this model to the idea of the spread of a misinformation item on Twitter, we applied new meaning to these terms. A person is Infected if they post a tweet about the misinformation item (in our case study, as tweet using the #DCblackout hashtag), and Susceptible if they have not yet posted any tweet. When a Susceptible individual comes into contact with an Infected individual via a tweet, that user will become Infected and will post a tweet about the target misinformation. In addition, Susceptible users remain in that susceptible state until they make contact with an Infected person [11].

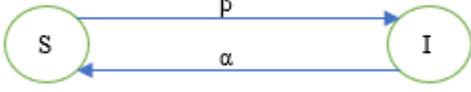

Fig. 1: SIS Model

The following system of Ordinary Differential Equations (ODE) represents the SIS model [11].

$$\frac{dS}{dt} = -\beta SI + \alpha I \quad (1)$$

$$\frac{dI}{dt} = \beta SI - \alpha I \quad (2)$$

### D. SIR Model

Another model, which is more practical and well known than the SIS model, is the SIR model. This model divides the population into three compartments (Fig. 2). The Susceptible (S) group consists of people who are in danger of contracting the disease but have not yet gotten the disease. The Infected (I) group consists of individuals who have the disease and are capable of spreading it to others. Finally, the Recovered (R) group consists of people who may not be Infected or cause anyone else to be infected because they either have Recovered from the disease and have become immune against it or they died from the disease [14].

To adapt this model toward the dissemination of misinformation on Twitter, we allocated new meaning to these terms. A person is Infected if they post a tweet about the misinformation item, and Susceptible if they have not yet posted any tweet but they are Exposed to the misinformation item and there is a possibility that they will subsequently post about the item, and Recovered if they have not subsequently posted about the misinformation item.

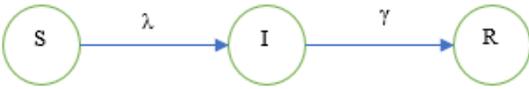

Fig. 2: SIR Model

The following system of Ordinary Differential Equations (ODE) represents the SIR model [14].

$$\frac{dS}{dt} = -\lambda S \quad (3)$$

$$\frac{dI}{dt} = \lambda S \quad (4)$$

$$\frac{dR}{dt} = \gamma I \quad (5)$$

### E. SEIZ Model

One major limitation of the SIS and SIR models is that when a Susceptible person is in contact with an Infected one, there is just one possibility, which is that the user can only move to the Infected component. This assumption does not apply well to the spread of misinformation, especially on social media, and specifically on Twitter.

Individuals can have different complicated beliefs when they are Exposed to items of misinformation on social media. Some people may have different viewpoints about the misinformation item; some others may need some time to come to believe it; or some others can be Skeptical to the accuracy of what they saw.

When people are faced with misinformation on social media, they may be convinced to spread the misinformation after thorough consideration, which requires some time for some people, while being immediate for others. In addition, there is a possibility that some individuals that are Exposed to an item of misinformation never show any reaction to it.

These possibilities are not covered by the SIS nor the SIR models. Based on the provided reasons, we decided to use the SEIZ model, which is a more powerful model and more applicable to the spread of misinformation on Twitter because, as we mentioned before, it includes the Skeptics component and Exposed component which is more suitable for the process of spreading misinformation.

In the context of analyzing the spread of misinformation on Twitter, the various components of the SEIZ model (Fig. 3) are outlined below.

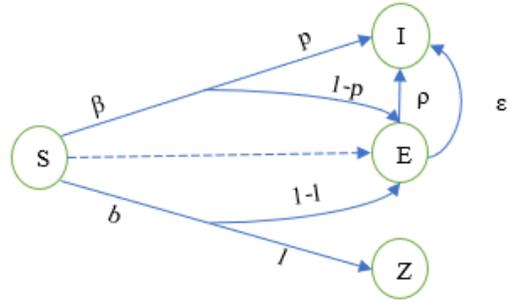

Fig. 3: SEIZ Model [11]

- Susceptible (S) represents individuals who have not yet heard about the misinformation.
- Exposed (E) represents the users who have been Exposed to the item of misinformation via a tweet and had a delay of time before posting a tweet about it themselves.
- Infected (I) relates to users who have tweeted about the misinformation item.

- Skeptic (Z) refers to individuals who have encountered the misinformation item via a tweet but chose not to tweet about it.

It is important to note that when we refer to the Z compartment as Skeptics it is not implying that the user actually has a set belief regarding that information item. We simply are adopting this terminology due to its use from the original authors of the SEIZ model [11].

One of the most important improvements of the SEIZ model over the SIR and SIS models lies in its seeming understanding that it is possible for people to encounter a misinformation item on twitter and not execute any reaction. Furthermore, in the SEIZ model, individuals can be Exposed to a misinformation item but not tweet about it immediately.

To be more precise, people have the opportunity to spend some deliberative time before deciding to believe it, which then allows that user to be transferred to the Infected component [11]. The following system of Ordinary Differential Equations (ODE) represents the SEIZ model [11]:

$$\frac{dS}{dt} = -\beta S \frac{I}{N} - bS \frac{Z}{N} \quad (6)$$

$$\frac{dE}{dt} = (1-p)\beta S \frac{I}{N} + (1-l)bS \frac{Z}{N} - \rho E \frac{I}{N} - \varepsilon E \quad (7)$$

$$\frac{dI}{dt} = p\beta S \frac{I}{N} + \rho E \frac{I}{N} + \varepsilon E \quad (8)$$

$$\frac{dZ}{dt} = lbS \frac{Z}{N} \quad (9)$$

For the above-mentioned ODEs, the parameters are defined in Table 1, and those parameters are explained in further detail below.

TABLE I
PARAMETER DEFINITIONS FOR THE SEIZ MODEL [11].

| Parameter | DEFINITION |
|---|---|
| β | Contact rate between S and I. |
| b | Contact rate between S and Z. |
| ρ | Contact rate between E and I. |
| p | Probability of S to I given contact with I. |
| 1-p | Probability of S to E given contact with I. |
| ε | Transition rate of E to I (Incubation rate). |
| l | Probability of S to Z given contact with Z. |
| 1-l | Probability of S to E given contact with Z. |

Susceptible (S) - The user has not yet heard about the misinformation item - comes into contact with Infected users (I) with the rate of β and can theoretically instantly believe an item of misinformation with the probability of p, or that individual may have some doubts and need some time to analyze the misinformation item when they have time and move to the Exposed (E) status with a corresponding probability of (1-p).

Skeptic (Z) - users who have heard about the misinformation item but choose not to tweet about it - recruited from the Susceptibles with the rate b. These activities can cause two different possibilities. The first one is that it can cause turning the user into another Skeptic with the probability l. This means that the user decides not to tweet about the misinformation item maybe because they don't believe it or believe it but decide not to pass it on. The second possibility is that it can cause the inadvertent result of sending the user into the Exposed (E) compartment with the probability (1-l).

There are two different possibilities for the transfer of users from the Exposed state to the Infected state. The first one is that people who are in the Exposed component - users who have heard about the misinformation item but need some time before posting it - may have more contact with Infected individuals who are people who post about the misinformation item with a contact rate ρ and because of this further contact they will become Infected. Another possibility is that users in the Exposed component can transfer to the Infected component because of self-adoption with rate ε rather than because of having more contact with Infected users. For example in our case study, users who have been Exposed to the hashtag #DCblackout may decide to post about it not immediately, but after some time that they need to make sure that it is true or maybe because they became aware of it from other platforms.

### F. Basic Reproduction Number

The potential of the contagiousness of a specific disease among a population depends on the "basic reproduction number" or ratio or rate, R0 (pronounced "R naught"). Although we are not assessing R0 in this work, we intend to incorporate it into our future work, and believe that it is important to provide a brief discussion. When an infectious person enters a susceptible crowd, the average number of people directly Infected by that person during their "contagious cycle" is defined as R0 [14], [28].

According to the "threshold theorem", if R0 is greater than 1, an epidemic happens because, on average, one infectious person is transmitting the infection to more than one susceptible person. If R0 is equal to 1, the infection stays endemic because one infectious person on average propagates the disease to just one other susceptible person. On the other hand, if R0 is smaller than 1, the infection finally vanishes from the population because, on average, one infectious person is transmitting the infection to fewer than one other susceptible person, and, consequently, the amplitude of the infection propagation would be smaller [29].

### G. Parameter Learning

Recall from equations 6 through 9, in order to apply the SEIZ model, we need to designate several parameters, including contact rates (β, b, ρ), probabilities (p, l), and a transition rate ε. In addition, in order to solve the ODEs we need to set initial values for each component in the model: $S(t_0)$, $E(t_0)$, $I(t_0)$, and $Z(t_0)$. These parameters are unknown, and, therefore, must be evaluated. Consequently, the total population size (N), which is the total number of users that exist in all four compartments is also unknown. The initial values for every component and its parameters are inputs for our model.

We utilized the *lsqnonlin* function, which is a nonlinear least square fit function in Matlab [34] and uses a trust-region-reflective algorithm, to fit the SEIZ model to our dataset. The ODEs were solved using *ode45*, which is a built-in function of Matlab. The set of parameters that minimizes the error between the actual number of tweets (in this case study, individuals who posted a tweet containing the #DCblackout hashtag) and the estimated number of users in the Infected compartment, |I(t) - tweets(t)| was identified as the optimal parameter set.

To produce the actual number of tweets in the Infected (I) component we used a cumulative total of tweets from the start of the spread of the misinformation item (June 1) based on every 15 minutes. Then we used the total number of tweets to determine the lower and upper bounds of the parameters we desired to optimize. First, we tried to calculate the parameters without these bounds. However, we understood that without upper and lower bounds, the model generates improbable results. Next, we present the analysis and results of the study.

## IV. ANALYSIS AND RESULTS

In this section the results of this study are presented. The first part is related to the fitting of our dataset to the Infected (I) compartment of the SEIZ model. Then we evaluated the transition of users through the other compartments--Susceptible (S), Exposed (E), and Skeptic (Z).

### A. Fitting Data To Infected Component

To fit our dataset with the SEIZ model, we first fit the number of Infected people (those users who used the hashtag #DCblackout) in each 15-minute time interval with the Infected (I) component in the SEIZ model using Matlab. As seen in Fig. 4, our dataset fits with the I component of the model very well, and the difference between the actual number of tweets and the Infected component is very low.

To be able to quantify this difference, we used relative error in 2-norm [11].

$$\frac{||I(t) - tweets(t)||_2}{||tweets(t)||_2} \quad (10)$$

The relative error in 2-norm for our dataset was very low (error = 0.019). As discussed in the previous section, we applied the SEIZ model to study the propagation of a specific misinformation item on Twitter. Again, to the best of our knowledge, no study has attempted to apply the SEIZ model to the spread of *misinformation* on social media. For comparison, using this technique, Jin et al. obtained a mean relative error of 0.074 while modeling the spread of *rumors* on Twitter [11].

Although our result may not be directly comparable (due to the previously discussed differences between the concepts of rumor and misinformation), our analysis is a crucial step towards understanding how misinformation spreads on a specific online social network, such as Twitter.

The low relative error obtained in our fitting the SEIZ model implies that this model correctly represents the Twitter data for an item of misinformation. So, when we have a mathematical model for the spreading of misinformation on Twitter, we can evaluate the number of individuals of any compartment at any time. In other words, using the mathematical model can help us to predict the number of Infected individuals (in this case study, users who posted the #DCblackout hashtag on Twitter) in the future.

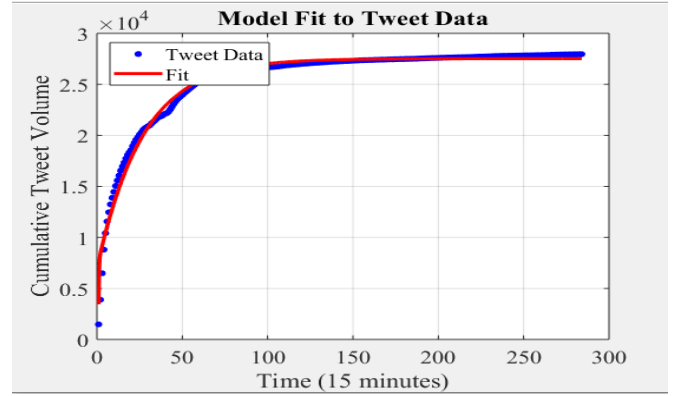

Fig. 4: Model fit to twitter data

### B. Analyzing Different Components

The next step of analysis that is now possible with the SEIZ model for our case study is the evaluation of the transition of individuals through the various compartments of the SEIZ model. Fig. 5 (a) illustrates this transition.

The parameters that produced the best fit are β = 4.3713, ρ = 1.3833e-06, and ε = 0.0373. Also, the probabilities are b = 8.1967, p = 0.7905, and l = 0.8161. These parameters are the Matlab output code which solves the ODEs and fits best to our dataset.

These parameters minimized the error between the actual number of tweets containing the #DCblackout hashtag and the estimated number of tweets in the Infected compartment of the SEIZ model.

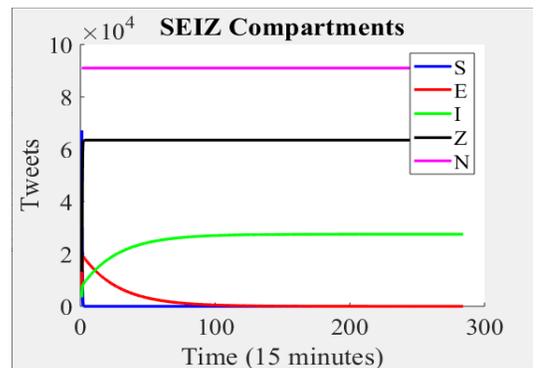

(a)

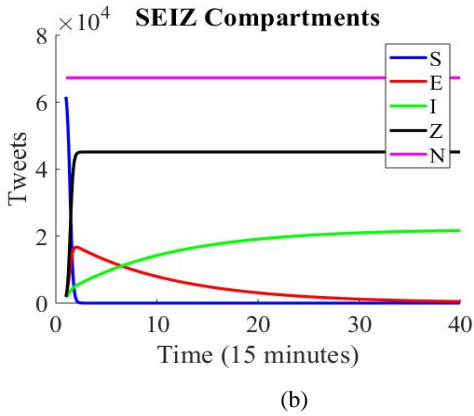

Fig. 5: Different components of the SEIZ model with parameters β = 4.3713, ρ = 1.3833e-06, and ε = 0.0373

As we can see in Fig. 5 (a), the number of users changes at the beginning of the time intervals and then remains constant. To better see the changes in the beginning of the process of the misinformation spread, we subsequently decided to limit our experiment to 40-time intervals (Fig. 5 (b)).

As we can see in Fig. 5(b), the number of Infected (I) individuals increased when the number of Susceptible (S) individuals decreased, but the rate of change in the Infected compartment is slower than the rate of change in the Susceptible compartment. Also, the greatest change in the Infected increments occurred when Susceptibles became stable to its minimal amount. This shows that the majority of users who became Infected did not directly transfer from the Susceptible compartment.

These findings indicate that the ratio of Susceptible users who transferred to the Infected compartment is much less than those that transferred to the Skeptic compartment. In addition, the increase in the Skeptic (Z) compartment occurred concurrently with the decrease in the Susceptible compartment (S). Consequently, both Susceptibles (S) and Skeptics (Z) became steady at the same time.

Moreover, we can see that the increase in the Exposed (E) compartment was strongly related to the increase in the Skeptic (Z) compartment, and both Exposed and Skeptics peaked at the same time. We can conclude that as the number of people who were doubtful about the accuracy of the misinformation item (and therefore needed some deliberative time before deciding to react) increased, the number of Skeptic users who decided not to tweet about the misinformation item also increased. Furthermore, users in the Exposed compartment began to decline at a negative rate and were symmetric to the rate of increase in the Infected compartment. Actually, the growth in the Infected (I) users directly coincided with a decrease in the Exposed (E) users. This means that almost all of the users who tweeted about the item of misinformation and became Infected (I) first needed some time to decide to believe and react, while a small number of them decided to immediately tweet about the misinformation item.

To summarize, we can conclude that all users are initially in the Susceptible compartment due to the nature of our dataset, which involves any Twitter user who was connected to the use of the #DCblackout hashtag. Due to contact with Skeptics, most of the Susceptible individuals transferred to the Skeptic compartment, and those Susceptible individuals who transferred to the Exposed compartment did so through their interaction with the Skeptic users. In other words, most of the people who viewed the misinformation item didn't have any reaction to it, while others needed some time before deciding to post about it after their interaction with the Skeptics. Moreover, the Infected users did not transfer directly from the Susceptible compartment but instead from the Exposed compartment.

In our study, the results showed that the incubation rate (ε) is 0.037 and the contact rate between E and I (ρ) is 1.3833e-06. These numbers show that the Exposed individuals transferred to the Infected compartment because of self-adoption and not so much because of directly coming into contact with Infected users. In other words, surprisingly, probably other mediums had more influence on a user's decision to spread an item of misinformation on Twitter than did their coming into contact with the tweets about that misinformation [11].

To summarize, the key findings of this study are:
- The majority of the Susceptible users, after viewing the misinformation item on Twitter, decided not to tweet about it.
- A very small number of Susceptible users, after viewing the misinformation item on Twitter, decided to post about it immediately.
- The majority of the people who became Infected and posted about the misinformation item transferred from the Exposed compartment, which implies that they needed some deliberative time before deciding to post about the misinformation.
- Most of the users who needed some time before deciding to post about the misinformation item, posted about it not because of direct contact with other Infected users, but instead most likely because they heard about the misinformation from another platform.

IV. CONCLUSION AND FUTURE WORK

In this paper, we showed how the spread of misinformation concerning the unrest in Washington, D.C. that was propagated on Twitter can be modeled by applying the SEIZ epidemiological model. Acquiring an error score of 0.019 between the real data from Twitter and the fitted model demonstrates that the SEIZ model is accurate in evaluating the spread of a particular misinformation item on Twitter.

Using this mathematical model for the spread of an item of misinformation can provide an opportunity to evaluate and forecast diffusion trends. This evaluation can help develop proper strategies for controlling the spread of misinformation on online social networks. In the future, we plan to apply the epidemiological model on more misinformation items and compare the results with those of the current work. Also, we plan to conduct a study on evaluating the spread of misinformation and real news and compare the diffusion trends. Applying the SEIZ epidemiological model to datasets from other social media platforms such as YouTube, Facebook, and others is also an intended research direction. We also intend to incorporate the important R0 metric into our future work using simulations, which will allow us to predict whether or not specific items of misinformation will ultimately become a "pandemic" in terms of virality and mobilization. Therefore, our future work also involves attempting to develop baseline parameters for allowing researchers to apply this model for prediction of misinformation across various social media platforms.

ACKNOWLEDGMENT

This research is funded in part by the U.S. National Science Foundation (OIA-1946391, OIA-1920920, IIS-1636933, ACI-1429160, and IIS-1110868), U.S. Office of Naval Research (N00014-10-1-0091, N00014-14-1-0489, N00014-15-P-1187, N00014-16-1-2016, N00014-16-1-2412, N00014-17-1-2675, N00014-17-1-2605, N68335-19-C-0359, N00014-19-1-2336, N68335-20-C-0540), U.S. Air Force Research Lab, U.S. Army Research Office (W911NF-17-S-0002, W911NF-16-1-0189), U.S. Defense Advanced Research Projects Agency (W31P4Q-17-C-0059), Arkansas Research Alliance, the Jerry L. Maulden/Entergy Endowment at the University of Arkansas at Little Rock, and the Australian Department of Defense Strategic Policy Grants Program (SPGP) (award number: 2020-106-094). Any opinions, findings, and conclusions or recommendations expressed in this material are those of the authors and do not necessarily reflect the views of the funding organizations. The researchers gratefully acknowledge the support.